\documentclass[aps,twocolumn,groupedaddress,showpacs,floatfix]{revtex4}

\usepackage{graphicx}
\usepackage{amsmath}

\begin{document}

\title{Trapping and cooling of rf-dressed atoms in a quadrupole magnetic field}
\author{O.~Morizot}
\author{C.\,L.~Garrido Alzar}
\author{P.-E.~Pottie}
\author{V.~ Lorent}
\author{H.~Perrin}\email{helene.perrin@galilee.univ-paris13.fr}

\affiliation{Laboratoire de physique des lasers, Institut Galil\'ee, Universit\'e Paris 13 and CNRS,
Avenue J.-B. Cl\'ement, F-93430 Villetaneuse, France }

\date{\today}

\begin{abstract}
We observe the spontaneous evaporation of atoms confined in a bubble-like rf-dressed trap~\cite{Zobay01}. The atoms are confined in a quadrupole magnetic trap and are dressed by a linearly polarized rf field. The evaporation is related to the presence of holes in the trap, at the positions where the rf coupling vanishes, due to its vectorial character. The final temperature results from a competition between residual heating and evaporation efficiency, which is controlled via the height of the holes with respect to the bottom of the trap. The experimental data are modeled by a Monte-Carlo simulation predicting a small increase in phase space density limited by the heating rate. This increase was within the phase space density determination uncertainty of the experiment.
\end{abstract}

%
%

\maketitle

\section{Introduction}
Ultracold atoms in versatile traps is a subject of extensive studies. Developing alternative trap geometries enables the exploration of new physical situations. Amongst the most remarkable results obtained with atoms in non harmonic traps, one may cite the observation of superfluid to Mott insulator transition in optical lattices~\cite{Greiner02}, Josephson oscillations in a double well~\cite{Albiez05} or trapping in a ring geometry~\cite{Gupta05}. Recently, radio-frequency (rf) fields were used together with a static magnetic field to produce a quasi 2D trap~\cite{Zobay01,Colombe04,White06} and a double well~\cite{Schumm05}. Other trapping geometries were proposed, based on this promising rf-dressing technique, such as rings~\cite{Morizot06,Lesanovsky06} or lattices~\cite{Courteille06}.
One presents here the implementation of rf-dressing inside a quadrupole magnetic trap.

Quadrupole traps are usually not popular for ultracold atom trapping due to Majorana losses at the centre, where the field vanishes. The losses are avoided in TOP traps by adding a rotating uniform magnetic field, at a frequency in the 10~kHz range~\cite{Petrich95}. In the trap presented in this paper, the magnetic field oscillates at a few MHz, resulting in an rf-dressed quadrupole trap. The atoms are then located away from the region of zero magnetic field. Still, the rf polarization is responsible for the existence of regions of zero rf coupling. This results in a leakage of the higher kinetic atomic population. In this Letter, we give evidence for a temperature decrease and a non exponential atom number loss, characteristic for an evaporation process, for different experimental parameters. 

\section{Trap description}
\label{sec:trap}
The trap relies on rf coupling between Zeeman sublevels in an inhomogeneous static magnetic field. The basic idea of the rf-dressed potentials was first proposed by Zobay and Garraway~\cite{Zobay01}, and experimentally demonstrated a few years later~\cite{Colombe04}. We will recall here the main features of this trap and develop further the consequences of an inhomogeneous rf coupling.

\begin{figure}
\includegraphics[width=0.7\columnwidth]{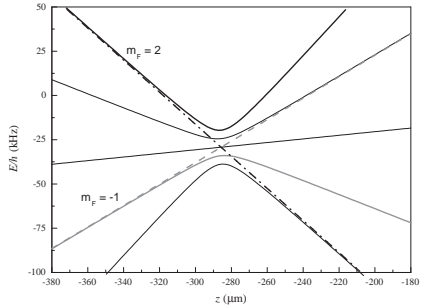}
\caption{Energy of the dressed levels of Rb in the magnetic quadrupole trap described in the paper, plotted along the vertical coordinate $z$, in the vicinity of the potential minimum at $z = -r_0$. The parameters are $\omega_{\rm rf} = 2\pi \times 3.1$~MHz, $b'=150$~G/cm and $\Omega_1=2\pi \times 100$~kHz. The five dressed sublevels for a $F=2$ spin state are plotted, as well as two bare states for comparison, $m_F = -1$, dashed, and $m_F = 2$, dash-dotted. The atoms are trapped in the upper dressed potential $m_F=F=2$. Gravity was taken into account.}
\label{fig:dressed_levels}
\end{figure}

A static trapping magnetic field $B(\mathbf{r}) \, \mathbf{e}_{\rm B}(\mathbf{r})$ presenting a local minimum $B_0$ is used together with a rf oscillating magnetic field $B_{\rm rf} \cos(\omega_{\rm rf} t)\,\mathbf{e}_{\rm y}$. The rf field is detuned by $\Delta=\omega_{\rm rf}-\omega_0$ from the coupling frequency $\omega_0=g_L \mu_B B_0/\hbar$ at the static magnetic field minimum, where $g_L$ is the Land\'e factor and $\mu_B$ the Bohr magneton. The linearly polarized rf field couples the Zeeman substates $|F,m_F\rangle$ and $|F,m'_F\rangle$ with $m'_F = m_F \pm 1$ at the positions where the magnetic splitting is close to the rf frequency. In the limit of a large coupling, this results in a dressing of the $m_F$ levels into adiabatic states whose energy are represented on fig.~\ref{fig:dressed_levels}. In the following, we will concentrate on the upper adiabatic state, undergoing a potential
\begin{equation}
V_d(\mathbf{r}) = F \left( (V_B(\mathbf{r}) - \hbar \omega_{\rm rf} )^2 + \hbar^2 \Omega^2(\mathbf{r}) \right)^{1/2} \, .
\end{equation}
This expression is obtained within the rotating wave approximation. Here, $m_F=F=2$ in the case of $^{87}$Rb in its $5s_{1/2}$ ground state. $V_B(\mathbf{r}) = g_L \mu_B B(\mathbf{r})$ is the magnetic energy shift between two adjacent Zeeman sublevels, and
\begin{equation}
\Omega(\mathbf{r}) = \Omega_1 \sin(\mathbf{e}_{\rm y},\mathbf{e}_{\rm B}(\mathbf{r}))
\end{equation}
is the Rabi frequency of rf coupling, of maximal amplitude $\Omega_1 = g_L \mu_B B_{\rm rf}/(2\hbar)$. $\Omega_1$ is independent of $\mathbf{r}$ to a very good approximation. As deduced from this formula, the rf coupling vanishes at the points where the static field is parallel to the radio frequency polarization $\mathbf{e}_{\rm y}$, that is along the $y$ axis. We will see later on that in the case of a quadrupole static magnetic field, these points appear as two ``holes'' on the equator of our bubble-like trap.

The simplest situation occurs when the rf coupling $\Omega(\mathbf{r})$ is uniform. Then, for a positive detuning $\Delta>0$, the dressed potential $V_d(\mathbf{r})$ is minimum all over an iso-$B$ surface, defined by $V_B(\mathbf{r}) = g_L \mu_B B(\mathbf{r}) = \hbar \omega_{\rm rf}$. This is no longer true if $\Omega$ varies in space, due to the spatial dependence of the orientation of the static field: this breaks the invariance on the iso-$B$ surface, leading to a finite number of minima related to the minima of $\Omega(\mathbf{r})$. On the other hand, gravity tends to push the atoms at the bottom of the iso-$B$ surface. To describe correctly the potential geometry, we thus need to take both effects into account. As we will see, the potential minima will not always sit exactly on the iso-$B$ surface, but they will be close to it in most relevant experimental cases.

\begin{figure}[t]
\begin{center}
\includegraphics[width=0.5\columnwidth]{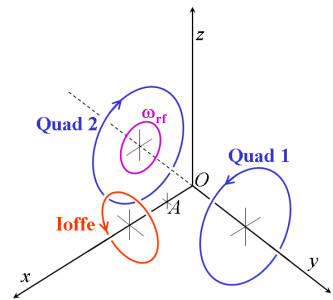}
\end{center}
\caption{Simplified scheme of the trap. After accomplishing evaporative cooling in the QUIC trap (Quad + Ioffe coils) centered at position $A$, the atoms are loaded into the dressed rf trap (Quad + Ioffe + rf coils). The current in the Ioffe coil is then slowly ramped down to zero for transferring the atoms into the dressed trap based on the quadrupole field of axis $y$ (Quad + rf coils). The rf field polarization is aligned along the axis of the quadrupole.}
\label{fig:coils}
\end{figure}

We concentrate in the following on the case of a quadrupole trap with an horizontal symmetry axis. This axis is labeled $y$ in our experimental setup, see fig.~\ref{fig:coils}, and is the same as the rf polarization. Let $b'$ be the magnetic field gradient in the radial direction of coordinate $\rho = (x^2 + z^2)^{1/2}$, and let us define $\alpha$ as $\alpha = g_L \mu_B b'/\hbar$. One has $\mathbf{B}(\mathbf{r}) = b'(\rho \mathbf{e}_{\rho} - 2 y \mathbf{e}_y)$ and therefore $\Omega(\mathbf{r}) = \Omega_1 \rho/\sqrt{\rho^2+4y^2}$. In this situation, an atom in the uppermost dressed state $m_F=F$ undergoes the potential
\begin{eqnarray}
&V_{\rm tot}&\!\!\!(x,y,z)= m g z \label{eqn:Vtot1}\\
&+& F \hbar\sqrt{ (\alpha \sqrt{\rho^2 + 4 y^2} - \omega_{\rm rf})^2 + \frac{\rho^2}{\rho^2 + 4 y^2} \, \Omega_1^2 } \nonumber
\end{eqnarray}
where $m$ is the atomic mass and $mgz$ the gravitational potential. The relevant iso-$B$ surface is then an ellipsoid of equation $\rho^2 + 4y^2 = r_0^2$, where $r_0$ is related to $\omega_{\rm rf}$ through $r_0 = \omega_{\rm rf}/\alpha$. The rf coupling vanishes along the $y$ axis, where the magnetic field is parallel to the rf polarization.

To visualize the potential, let us first remark that the extrema are located in the vertical plane $x=0$. In this plane, $\rho=|z|$. One can rewrite $V_{\rm tot}(y,z)$ in units of $mgr_0$ as:
\begin{eqnarray}
&&V_{\rm tot}(x=0,y,z)/ (m g r_0) = \\[2mm]
&&\frac{z}{r_0} +\beta\sqrt{ \left(\frac{\sqrt{z^2 + 4 y^2}}{r_0} - 1\right)^2 + \frac{z^2}{z^2 + 4 y^2} \, \frac{\Omega_1^2}{\omega_{\rm rf}^2} }\nonumber \, .
\end{eqnarray}
The parameter $\beta=F \hbar \alpha/mg$ is the ratio of the magnetic and the gravitational force. It should be larger than 1 for the magnetic trap to compensate gravity and in our experiment $\beta\simeq10$. Therefore, the behaviour of the second term in the potential expression is dominant. Indeed, the exact study of the minima position of $V_{\rm tot}$ shows than for $\beta \gg 1$, the minima lie very close to the iso-B surface $\rho^2 + 4y^2 = r_0^2$. To discuss qualitatively the potential geometry, we will thus assume that the atoms sit exactly on this ellipsoid, which cancels the first term under the square root. We are left with:
\begin{equation}
V_{\rm tot}(z)/ (m g r_0) \simeq \frac{z}{r_0} +\beta\frac{|z|}{r_0} \, \frac{\Omega_1}{\omega_{\rm rf}}, \mbox{ with } |z|\leq r_0 \, .
\end{equation}

\begin{figure*}[t]
\begin{center}
\includegraphics[width=40mm]{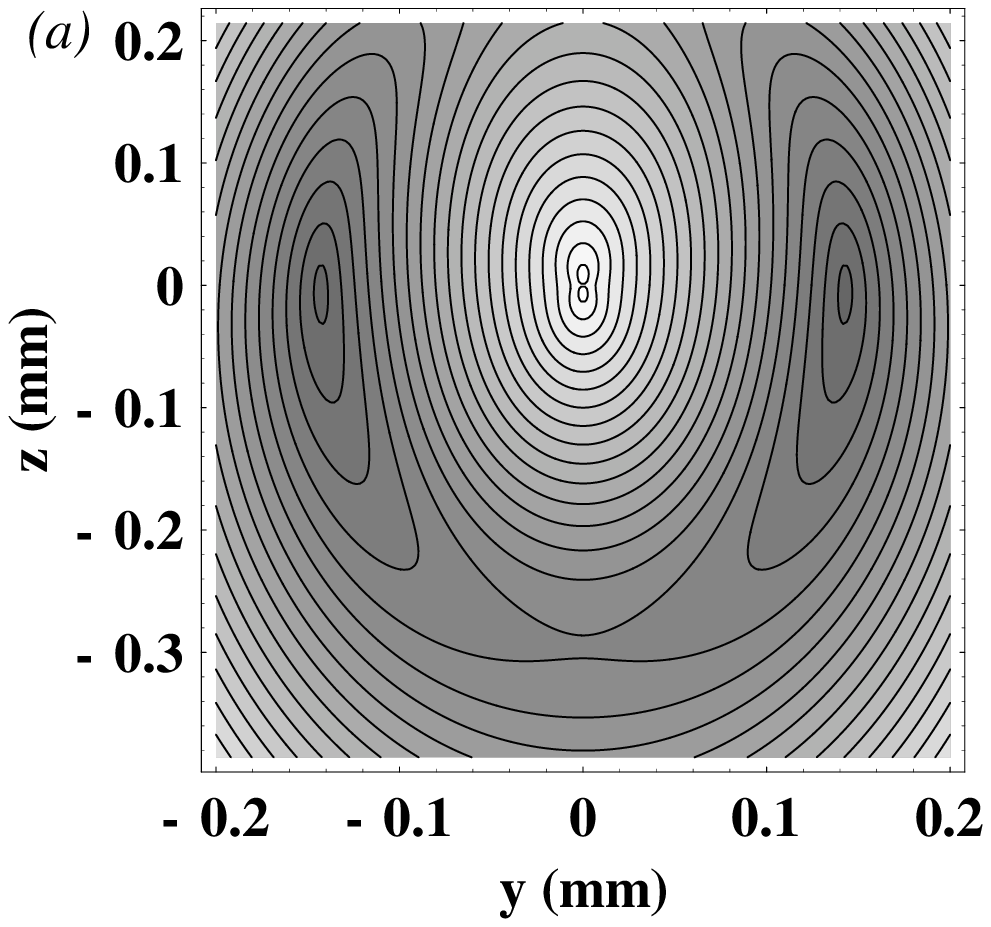} \hspace{2mm} \includegraphics[width=40mm]{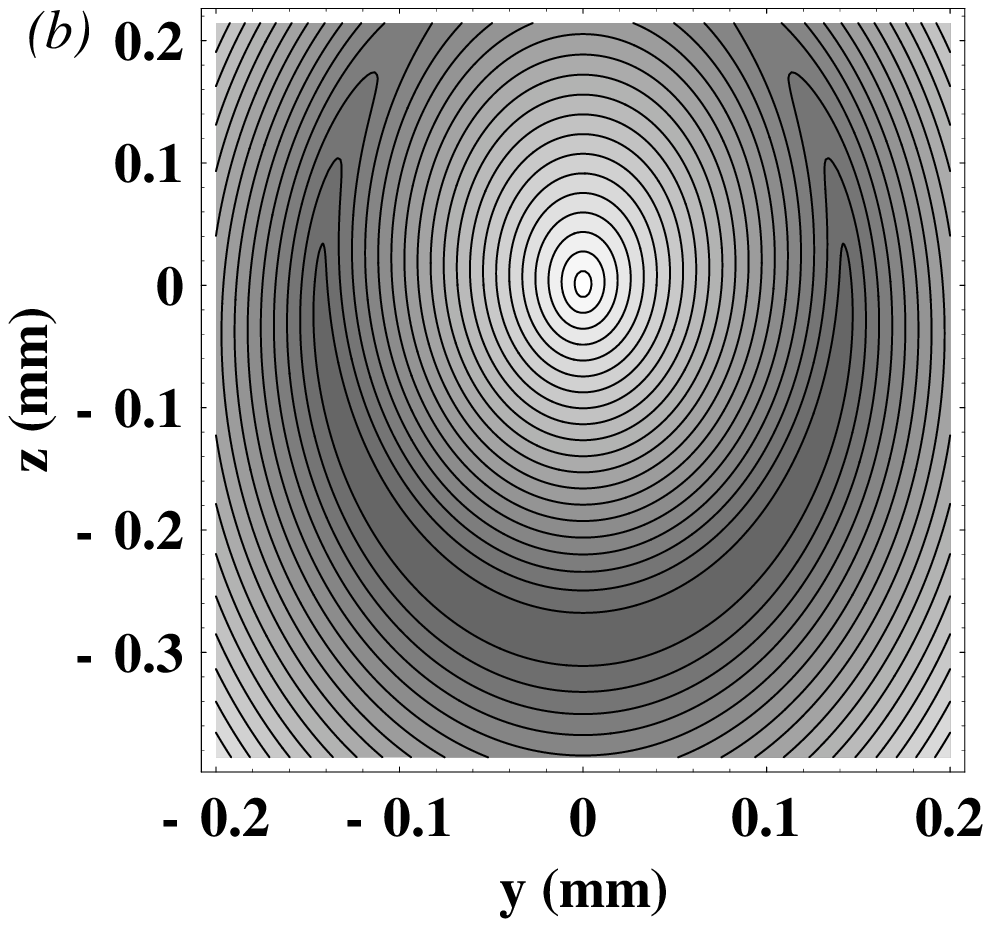} \hspace{2mm}
\includegraphics[width=40mm]{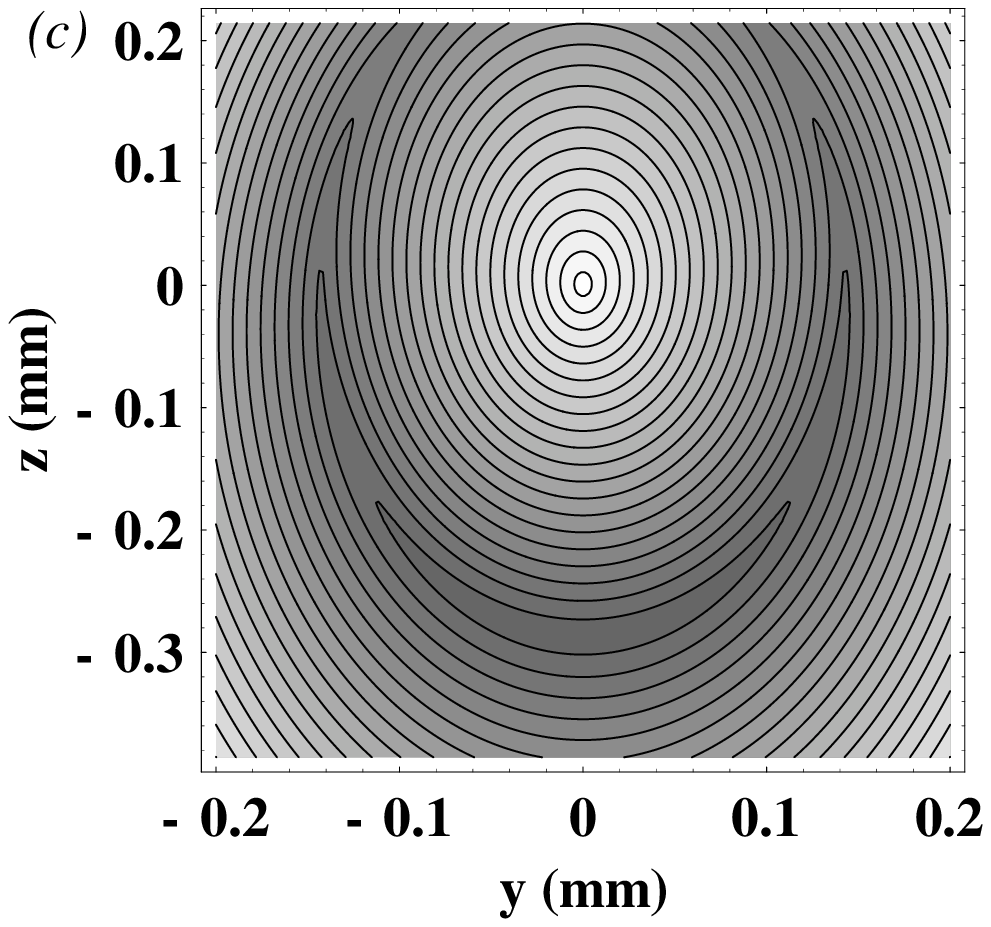}
\end{center}
\caption{Contour plot of the potential for Rb atoms in the $yz$ plane, for different values of the rf coupling relative to $\omega_{\rm rf}$. The parameters are $\omega_{\rm rf}/2\pi = 3.1$~MHz and a magnetic gradient of 150~G/cm in the $z$ direction, such that $\beta = 9.85$ and $\omega_{\rm rf}/\beta = 2\pi \times 305$~kHz. $(a)$ large coupling $\omega_{\rm rf}/\Omega_1 < \beta$ ($\Omega_1/2\pi=1$~MHz); the two minima are located at the position of the holes. $(b)$ intermediate coupling $\omega_{\rm rf}/\Omega_1 = \beta$ ($\Omega_1/2\pi=305$~kHz). $(c)$ small coupling $\omega_{\rm rf}/\Omega_1 > \beta$ ($\Omega_1/2\pi=100$~kHz); a single minimum is present, at the bottom of the shell.}
\label{fig:yzcut}
\end{figure*}

Depending on the ratio $\omega_{\rm rf}/\Omega_1$ of the rf frequency to the rf coupling, two different situations may occur: for $\omega_{\rm rf}/\Omega_1 < \beta$, or equivalently $m g r_0 < F\hbar \Omega_1$, the coupling inhomogeneity is the dominant effect, and the potential minima sit at the points where the coupling term is the smallest. This corresponds to the positions where $\rho = 0$, that is at the intersections of the iso-B ellipsoid and the $y$ axis (see fig.~\ref{fig:yzcut}$a$). Note that at these points, the rf coupling vanishes, leading to spin flip losses into untrapped states. In the following, we will refer to these two points as to the ``holes''. On the other hand, for $\omega_{\rm rf}/\Omega_1 > \beta$, or $m g r_0 > F\hbar \Omega_1$, gravity dominates and there is a single potential minimum at the very bottom of the ellipsoid, far from the holes (fig.~\ref{fig:yzcut}$c$). In this single well, and in the harmonic approximation, the oscillation frequencies up to first order in $1/\beta$~\cite{note1} are:
\begin{eqnarray}
\omega_x &=& \sqrt{\frac{g}{r_0}} \label{eq:nux} \, .\\
\omega_y &=& 2\sqrt{\frac{g}{r_0}}\sqrt{1-\frac{F\hbar\Omega_1}{m g r_0}} = 2\omega_x\sqrt{1-\beta\frac{\Omega_1}{\omega_{\rm rf}}} \, . \label{eq:nuy}\\
\omega_z &=& \alpha \sqrt{\frac{F \hbar}{m\Omega_1}} = \omega_x\sqrt{\beta\frac{\omega_{\rm rf}}{\Omega_1}} \, . \label{eq:nuz}
\end{eqnarray}
The vertical oscillation frequency was derived in \cite{Zobay01}. One easily identifies the pendulum frequency along $x$. Along $y$, the pendulum frequency is modified by the inhomogeneous coupling. We remark that these oscillation frequencies are valid only for small amplitude oscillations around the potential minimum, the trap as a whole being far from harmonic.

For values of $\omega_{\rm rf}/\Omega_1$ close to $\beta$, one will face a situation where the potential is almost flat in the crescent defined by $z<0$, $x=0$ and $z^2+4y^2=r_0^2$ (fig.~\ref{fig:yzcut}$b$). The gravitational energy $m g r_0$ at the holes is indeed exactly equal to the energy shift $F \hbar \Omega_1$ induced by the rf at the bottom of the iso-$B$ surface, and $\omega_y$ as expressed above goes to zero. However, this intermediate region is extremely narrow for $\beta \gg 1$, with a width in the parameter $\omega_{\rm rf}/\Omega_1$ on the order of $3/(2\beta)$ around $\beta$, and in practice one deals only with one of the two extreme situations, $a$ or $c$. Note that in any case the zero magnetic field region, responsible for Majorana spin flips in the centre of conventional quadrupole traps, plays a minor role in an rf-dressed quadrupole trap. The energy barrier is at least $F\hbar \omega_{\rm rf}$, much larger than the thermal energy of the atomic cloud.

In our experiment, $\omega_{\rm rf}/\Omega_1$ is larger than $\beta$ (typically $\omega_{\rm rf}/2\pi\simeq3$~MHz and $\Omega_1/2\pi < 260$~kHz), and we are always in the case of a single minimum. Therefore, due to gravity, ultracold atoms remain trapped near the bottom of the ellipsoid and only marginally explore the non-coupling regions. This single-well situation was demonstrated experimentally in a Ioffe-Pritchard type trap~\cite{Colombe04}. The opposite situation was used in an atom chip experiment for producing a double well potential~\cite{Schumm05}. In the experiment described in this paper, we bring into light the non negligible effect of the two holes for an ultracold atomic cloud, even in the situation where a single minimum is present.

\section{Experimental sequence}
The experimental setup has been described in detail elsewhere \cite{Colombe03,Dimova06}. The atoms in the state $F = 2$, $m_F = +2$ are first loaded into a static magnetic trap, in a Quadrupole and Ioffe Configuration (QUIC)~\cite{Esslinger98}, see fig.~\ref{fig:coils}, Quad and Ioffe coils. In this trap, the atoms are confined near the magnetic field minimum, position $A$ of fig.~\ref{fig:coils}, situated 7~mm away from the quadrupole center $O$ in the direction of the Ioffe coil. Then, a 30~s radio frequency ramp is applied to evaporatively cool them down to just above the condensation threshold. We deliberately do not reach condensation and rather work with ultracold clouds, with a temperature of typically $4~\mu$K, in order to explore the leakage through the pierced part of the iso-B bubble as explained later. The evaporation rf source is then switched off and the trapping rf field, polarized along $y$, is turned on at a frequency lower than the resonance frequency in the trap centre, 1.3~MHz. Note that for this purpose, a non zero minimum magnetic trap is necessary. The rf frequency is then adiabatically ramped up to the desired value of $\omega_{\rm rf}$ within typically 150~ms, as detailed in \cite{Colombe04}. This way, the atoms are always following the upper dressed state, corresponding to $m_F = F = +2$ in the dressed basis. At the end of this step, they are confined in a dressed rf trap relying on the QUIC static magnetic field.

The QUIC trap stage is necessary to obtain an ultracold sample and load it efficiently into the dressed rf trap. The atoms are then transferred into a dressed quadrupole trap, which presents points of strictly zero rf coupling at any value of $\omega_{\rm rf}$. The current in the Ioffe coil is therefore decreased in order to reach a quadrupole configuration within a time interval of about 450~ms. During this process, the minimum of the static magnetic field quickly goes to zero and is split into two zero field points separating along the $x$ axis. One point is going back 7~mm away from $A$ to the position $O$ of the initial quadrupole centre whereas the other one is going from $A$ to infinity in the opposite direction (see Ref.~\cite{Esslinger98}, Fig.~2). During this transfer, the atomic cloud remains in the dressed trap, situated just below the static magnetic minimum, and is split into two clouds according to the deformation of the iso-magnetic surfaces. The first stage of the Ioffe current decreasing is controlled carefully in order that at least half of the atoms follow the right direction, toward the initial quadrupole centre $O$. At the end of this stage, about $4\times10^5$ atoms at a higher temperature of $8~\mu$K are confined in an rf-dressed quadrupole trap. The QUIC-to-quadrupole transfer of the rf-dressed atoms is a critical step and is very sensitive to initial conditions. This procedure is at the origin of a scattering in the values of atom number and temperature.

The rf field is produced by a home made synthesizer~\cite{Morizot07}. The Rabi frequency $\Omega_1$ corresponding to a given voltage amplitude is evaluated as follows. The rf antenna, normally dedicated to rf trapping, is used for evaporative cooling in the QUIC trap for this measurement. An evaporation ramp is applied down to a given final rf frequency. The threshold frequency below which all atoms are expelled from the magnetic trap is recorded. For an arbitrary weak rf power, this threshold frequency corresponds to the Zeeman splitting $\omega_0$ at the bottom of the QUIC magnetic trap. However, the recorded threshold frequency is larger, as the magnetic levels are deformed due to dressing by the rf photons. We use the observed value as input in the calculation of the dressed potential, including gravity, and search for the Rabi frequency value making the potential flat, just unable to hold the atoms. We repeat this procedure for different voltage amplitudes at the synthesizer to calibrate the rf power.

\section{Experimental results}
\begin{figure}
\includegraphics[width=0.7\columnwidth]{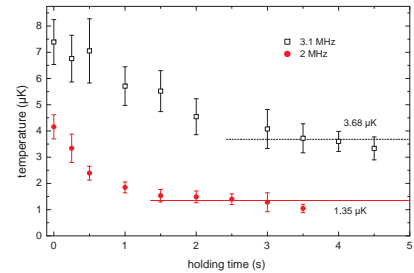}
\caption{Temperature $T$, measured as described in the text, as a function of time with $\Omega_1=2\pi \times 40$~kHz and for different values of the RF frequency $\omega_{\rm rf}/2\pi = 2$~MHz (closed circles) and $3.1$~MHz (open squares). Each point is an average over 2 to 3 data points. The error bars take into account the point dispersion as well as an experimental uncertainty of 10\%. $T$ drops over the first trapping seconds, and then stabilizes to a value $T=1.35~\mu$K (resp. $3.68~\mu$K) indicated by the horizontal lines, deduced from the average of the 5 (resp. 4) last points.}
\label{fig:2and3MHz}
\end{figure}

After the atoms are transferred to the rf-dressed quadrupole trap, they remain stored for a variable time $\tau$ after which the magnetic field and rf coupling are simultaneously switched off. An absorption image of the cloud is taken after a 7.5~ms ballistic expansion. From these measurements, the evolution of the atom number and the temperature with the storage time $\tau$ is deduced. The temperature is related to the vertical size after expansion of the atomic cloud. In principle, it also depends on the initial vertical size, which is not entirely negligible due to the curved shape of the trap. However, we neglect the initial size, which overestimates the real temperature by about 20\%, as deduced from the comparison with a numerical simulation, see next section. 

We observe a rapid decreasing of the temperature during the first few seconds. The experiment was repeated at $\Omega_1/2\pi=40$~kHz for two different values of $\omega_{\rm rf}$ (see fig.~\ref{fig:2and3MHz}). For $\omega_{\rm rf}/2\pi=2$~MHz, we find a final temperature $T_1=1.35~\mu$K whereas it does not get lower than $T_2=3.68~\mu$K for $\omega_{\rm rf}/2\pi=3.1$~MHz. For the largest rf coupling value $\Omega_1/2\pi=260$~kHz, the temperature decrease, can be observed over more than ten seconds (see the experimental points on fig.~\ref{fig:temperature}), with a first drop in temperature followed by a transition to a slower but still decreasing evolution.

These results, together with the non exponential decrease of the atom number, see inset of fig.~\ref{fig:temperature}, reveal an evaporation process via Landau-Zener losses through the holes along the rf coil axis in the equatorial plane. The final temperature is related to the trap depth, calculated between the trap bottom and the position of the holes and depending on the rf frequency through
\begin{equation}
U_0 = m g r_0 - F \hbar \Omega_1 = \frac{m g \omega_{\rm rf}}{\alpha} - F \hbar \Omega_1 \,.
\end{equation}
The trap depth is $U_0/k_B=\Theta_1=26.7~\mu$K (resp. $\Theta_2 = 15.7~\mu$K) for a 3.1~MHz (resp. 2~MHz) rf frequency and a Rabi frequency $\Omega_1/2\pi=40$~kHz. The final temperature compared to the trap depth gives $\Theta_1/T_1=7.3$ and $\Theta_2/T_2=11.6$. The overestimation of temperature makes these values a lower bound to the depth to temperature ratio. A ratio of 8 to 10 is considered typical for a spontaneous evaporation process in an harmonic trap~\cite{Luiten96}.

In principle, the evaporation could also come from losses through the centre of the quadrupole, however the energy $F\hbar \alpha r_0 = F\hbar\omega_{\rm rf}$ required to cross the central region of the quadrupole trap is much higher than the energy $U_0$ necessary to explore the uncoupled region, making this process very unlikely.

Finally, the possible non-adiabatic Landau-Zener losses in the avoided crossing region at the bottom of the ellipsoid, where the atoms mostly sit, can be neglected. Indeed, the probability of spin-flipping in this zone for a given velocity $v$ reads \cite{Suominen}
\begin{equation}
P_{LZ} = 1-\left[1 - \exp\left(-\frac{\pi}{2}\frac{\Omega_1^2}{\alpha v}\right)\right]^{2F} .
\label{LZ_losses}
\end{equation}
If one averages Eq. (\ref{LZ_losses}) over the whole velocity distribution for an rf coupling $\Omega_1 = 30$~kHz, smaller than was ever used in the experiment, and a temperature in the dressed trap $T = 10~\mu$K, one finds a life time $\tau_{LZ}$ due to Landau-Zener losses as high as 100~s, which is even larger than the life time due to collisions with the background gas. Landau-Zener losses at the trap bottom are then obviously not responsible for the losses and temperature decrease.

\section{Numerical simulation}
\begin{figure}
\includegraphics[width=0.7\columnwidth]{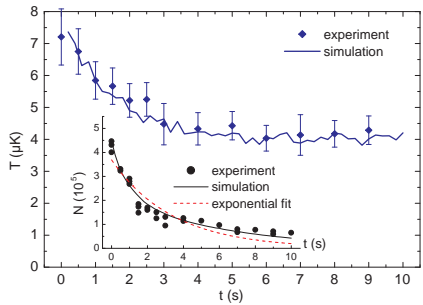}
\caption{Temperature, deduced from the size $\sigma_{z}$ after a 7.5~ms time-of-flight, as a function of the holding time in the rf-dressed quadrupole trap. Experimental data, closed diamonds with error bars as in Fig.~\ref{fig:2and3MHz}, are compared to the numerical simulation, thick line. Inset: Evolution of the atom number in the same conditions: experimental data, circles, are compared to the simulation, line and to a pure exponential fit, dashed red line. For these experiments and the related simulation, the parameters are $\omega_{\rm rf}/2\pi=3.1$~MHz and $\Omega_1=2\pi \times 260$~kHz.}
\label{fig:temperature}
\end{figure}

In order to confirm our interpretation, we performed a numerical simulation of the cloud classical dynamics with the parameters of the experiment, for a coupling $\Omega_1/2\pi=260$~kHz, a rf frequency $\omega_{\rm rf}/2\pi = 3.1$~MHz and a trapping time ranging between 0 and 10~s. The simulation calculates the evolution of a cloud of atoms initially trapped on the iso-B surface, at temperature $T_i$, above condensation threshold. Collisions between trapped atoms are taken into account through a method developed by G.A. Bird \cite{Bird}, while collisions with the background gas are described through a lifetime of 60~s. A possible linear heating rate $H$ is taken into account in the simulation by increasing each velocity component $v_i$ by $2 v_i H dt/T$ after a time step $dt$. The Landau-Zener losses are more difficult to model. The use of a random number to decide, with a local probability distribution, if an atom is lost or not would be time-consuming. Instead, the holes are modeled by a deterministic loss term: the atoms are considered lost if the Landau-Zener probability at this point $P_{LZ}(\mathbf{r},v)$ exceeds a critical value $P_c$. Due to the exponential dependence of the Landau-Zener non adiabatic coupling on the rf Rabi frequency, this occurs essentially in a region very close to the two holes.

Fig.~\ref{fig:temperature} gives a plot of the temperature for different holding times in the dressed trap. The temperature is deduced in the same conditions as in the experiment, from the cloud size after a 7.5~ms time-of-flight, and the two values are thus directly comparable. The simulation fits at best the experimental data for the choice $P_c=0.9$ and $H=300~$nK/s. The Landau-Zener probability $P_c$ mainly determines the initial rate of temperature decrease. Once this value is fixed, the heating rate $H$ imposes the final saturation temperature, resulting from the equilibrium between evaporation and residual heating. The simulation reproduces well the features of spontaneous evaporation, namely temperature reduction and non exponential atomic losses, see dotted line in the inset of fig.~\ref{fig:temperature}. These features completely disappear if the holes are suppressed. 

\begin{figure}
\includegraphics[width=0.7\columnwidth]{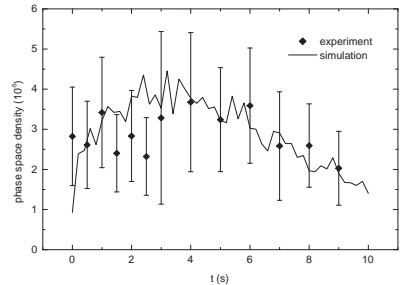}
\caption{Estimated phase space density for the data series of Fig.~\ref{fig:temperature}, from the experimental data, diamonds, and the simulation, line. The parameters are $\omega_{\rm rf}/2\pi=3.1$~MHz and $\Omega_1=2\pi \times 260$~kHz. The error bars take into account data dispersion and an experimental uncertainty of 40\%, resulting from uncertainties in atom number and temperature.}
\label{fig:psd}
\end{figure}

The peak phase space density $\varphi$ in the rf-dressed trap is computed from the simulated data. To be able to compare this figure with the experiment, we used in both cases the estimation $\varphi \simeq N (\hbar \bar{\omega}/k_B T)^3$, where $T$ is calculated from the vertical width after time-of-flight as explained above. Here, $\bar{\omega}$ is the geometric mean of the oscillation frequencies of the rf-dressed trap, Eq.(\ref{eq:nux}-\ref{eq:nuz}). This estimate only gives the order of magnitude for the phase space density, as the formula is correct only for a harmonic trap and the trap is far from being harmonic. The results are displayed on Fig.~\ref{fig:psd}, for the same parameters as used in Fig.~\ref{fig:temperature}. For these parameters, $\bar{\omega}/2\pi=60.6$~Hz. The simulation shows a first increase of about 50~\% within the first 3 seconds, followed by a regular decrease due to residual heating. The experimental data give values for $\varphi$ of the same order, but are too scattered for demonstrating unambiguously a phase space density increase. This is due to the sensitivity of the estimate for $\varphi$ to the temperature value $T$, which is determined with a large uncertainty.

\section{Conclusion}
In this paper, we demonstrate trapping of ultracold atoms in a quadrupole magnetic configuration. Atoms are prevented to escape through the centre by rf-dressing of the atomic state, which results in a confining adiabatic potential. Non exponential losses associated to a temperature decrease are observed in this trap, giving evidence for spontaneous evaporation. These losses are attributed to the presence of zero rf-coupling zones in the trap, due to rf polarization. Experimental data are interpreted with the help of a numerical simulation, well reproducing the main features. The evaporation results in a drop in temperature, which saturates due to a residual heating in the trap. This technical heating clearly prevents an increase in phase space density. Cooling in the trap may be improved in different ways. The observed spontaneous evaporation could be forced by reducing the rf frequency $\omega_{\rm rf}$ with time, thus limiting the trap depth. Still, it seems that a necessary approach would be to use a second independent rf field to spin flip selectively the more energetic atoms, as in conventional evaporative cooling~\cite{Garrido06,Hofferberth06}. Ultimately, the quadrupole geometry with a symmetry axis oriented in the vertical direction is ideal for the realization of a isotropic quasi-2D degenerate gas~\cite{Pitaevskii} or an atomic ring~\cite{Morizot06}.

\acknowledgments We acknowledge fruitful discussions with Barry Garraway. This work was supported by the R\'egion Ile-de-France (contract number E1213) and by the European Community through the Research Training Network ``FASTNet'' under contract No. HPRN-CT-2002-00304 and Marie Curie Training Network ``Atom Chips'' under contract No. MRTN-CT-2003-505032. Laboratoire de physique des lasers is UMR 7538 of CNRS and Paris 13 University. The LPL group is a member of the Institut Francilien de Recherche sur les Atomes Froids.


\begin{thebibliography}{99}
\bibitem{note1} The exact value of the oscillation frequencies contains other terms of orders at least $1/\beta^2$. They can be calculated analytically, but are negligible in our experimental conditions where $\beta \simeq 10$.
\end{thebibliography}
\end{document}